\documentclass[12pt]{iopart}
\usepackage{bm}
\usepackage{graphicx}

\begin{document}

\title{Linear amplification and quantum cloning for non-Gaussian continuous variables}

\author{Hyunchul Nha$^{1,*}$, G.~J. Milburn$^{2}$, and H.~J.~Carmichael$^{3}$}
\address{$^1$Department of Physics, Texas A \& M University at Qatar, POBox 23874, Doha, Qatar\\$^2$Centre for Quantum Computer Technology, School of Mathematics and Physics,
The University of Queensland, QLD 4072 Australia.\\
$^3$Department of Physics, University of Auckland,  Private Bag 92019, New Zealand}
\ead{hyunchul.nha@qatar.tamu.edu}

\begin{abstract}
We investigate phase-insensitive linear amplification at the quantum
limit for single- and two-mode states and show that there exists
a broad class of non-Gaussian states whose nonclassicality survives
even at an arbitrarily large gain. We identify the corresponding
observable nonclassical effects and find that they include, remarkably,
two-mode entanglement. The implications of our results for quantum
cloning outside the Gaussian regime are also addressed.
\end{abstract}
\maketitle

\section{Introduction}

When one amplifies a quantum state, a certain amount of noise is introduced as a fundamental requirement with profound implications for quantum communication and measurement; noise disallows the generation of perfect quantum clones \cite{Wooters}, which enables secure quantum communication, and it requires that quantum measurement---which necessarily involves amplification---be accompanied by decoherence, disallowing the mapping of a microscopic pure state to a macroscopic pure state.

Following the work of Caves \cite{Caves}, which established quantum-noise limits for both phase-sensitive and -insensitive linear amplification, numerous studies have investigated how phase-insensitive linear amplification (PILA) washes out nonclassical effects. Hong {\it et al.} \cite{Hong}, in particular, showed that sub-Poissonian statistics and squeezing disappear below the quantum cloning limit (gain $G=2$), and it has been argued that nonclassical effects disappear completely at sufficiently large gain \cite{Mandel}. The latter argument relies, however, on {\it non-optimal} amplification ({\it e.g.}, by a bath of two-state atoms \cite{note}); in quantum-limited amplification, states can remain nonclassical at arbitrarily large gain---{\it e.g.}, this is known for Fock states \cite{Mollow} and Schr{\"o}dinger-cat states \cite{Huang}. It is known also that amplitude-squared squeezing \cite{Hillery} and certain higher-order photon statistics \cite{Agarwal} can persist slightly above $G=2$. More generally, which nonclassical states or nonclassical effects persist beyond the cloning limit has not yet been systematically studied. Moreover, the study of PILA has been confined mostly to single-mode states, and there are a number of crucial issues to address outside the single-mode regime; in particular, the question of whether entanglement, or nonclassical correlation, can survive is of broad interest to quantum information science \cite{Yu}.

In addition to these consideration of PILA {\it per se\/}, linear amplification is generally the basis of quantum cloning schemes, particularly for continuous variables (CVs). A quantum cloner based on PILA and a beam-splitter has been identified as the optimal Gaussian cloner for coherent states \cite{Cerf1}, and the scheme can be extended to other CV states as a general tool for quantum cloning. Little is known, however, about the behavior of non-Gaussian states under linear amplification. It is thus important to carry the study of PILA systematically beyond the treatment of the Gaussian regime found in earlier works \cite{Cerf1,Agarwal1}.

In this paper we investigate both single- and two-mode states under
the nonunitary (phase-insensitive) process of PILA; our treatment
naturally extends to multi-mode states. As the unitary
(phase-sensitive) amplifier is able to {\it create} nonclassicality,
{\it e.g.}, squeezing, it is ruled out in this work. We connect the
critical gain at which all nonclassical effects disappear to the
nonclassical depth defined by Lee \cite{Lee1}, and identify effects
that remain observable at large gain. We show that there exists a
broad class of single- and two-mode non-Gaussian states whose
nonclassicality, remarkably including two-mode entanglement,
survives even at arbitrarily large gain, in sharp contrast to the
behavior known for Gaussian states \cite{Agarwal1}. This might also
be contrasted with the recent work by Allegra {\it et al.} that
presents evidence in support of the maximum robustness of Gaussian
entangled states against noisy thermal environments \cite{Paris}.
Furthermore, we show that the PILA-based quantum cloner can make
more than two clones in a nonclassical state with fidelity better
than classical scheme if the input has nonclassical depth larger
than $1/2$. Other aspects of PILA-based quantum cloning are also
briefly addressed.

\section{Phase insensitive linear amplification}

We begin with some background to quantum-limited PILA. Consider a mode with creation and annihilation operators $\hat a^\dag$ and $\hat a$, which after amplification at gain $G$ yields
\begin{eqnarray}
\hat a_G=\sqrt{G}\hat a+\sqrt{G-1}\hat v^\dag.
\label{eqn:PILA}
\end{eqnarray}
In Eq.~(\ref{eqn:PILA}) $\hat v$ is a vacuum mode introduced to account for the fundamental noise in the amplification; $\hat a_G$ thereby becomes a legitimate field operator, with commutator $[\hat a_G,\hat a_G^\dag]=1$. After taking the trace over the mode $v$, the action of the amplifier can be described by a completely positive map, with Kraus decomposition
\begin{eqnarray}
\rho_{\rm out}={\cal A }[\rho_{\rm in}]=\sum_{n=0}^\infty \hat{A}_n\rho_{\rm in}\hat{A}_n^\dagger,
\label{eqn:Milburn}
\end{eqnarray}
where
\begin{eqnarray}
\hat{A}_n=(G-1)^{n/2}\exp[-{\rm ln}(G)\hat a\hat a^\dagger/2](\hat a^\dagger)^n/\sqrt{n!}\mkern3mu.
\end{eqnarray}

In practice, quantum-limited PILA can be achieved by injecting the signal mode~$a$ into a nondegenerate parametric amplifier (NDPA) whose idler is in the vacuum state. Remarkably, as in \cite{Josse, Andersen}, it can also be realized using only linear optics by employing homodyne detection followed by feed-forward. In this paper, the NDPA framework is mostly employed. Thus, the two-mode input to the amplifier is written $\rho_{av}=\rho_{\rm in}\otimes|0\rangle\langle0|$, where $\rho_{\rm in}=\int d^2\alpha P_{\rm in}(\alpha)|\alpha\rangle\langle\alpha|$ is the signal-mode input to be amplified, with $P_{\rm in}(\alpha)$ being its $P$-function.
The amplified output is given by $\rho_G={\rm tr}_v(\hat U_{\rm PA}\rho_{av}\hat U_{\rm PA}^\dag)$, where $U_{\rm PA}\equiv \exp[\kappa t (a^\dag v^\dag-av)]$ executes amplification with gain $G\equiv\cosh^2(\kappa t)$.
After some calculation, the $P$-function of the amplified output is found to be 
\begin{eqnarray}
P_G(\alpha)=\frac{1}{\pi(G-1)}\int\mkern-3mu d^2\beta P_{\rm in}(\beta)\exp\mkern-3mu\left(-\frac{\left|\beta-\alpha/\sqrt G\mkern2mu\right|^2}{1-1/G}\right)\mkern-3mu,
\label{eqn:P-trans}
\end{eqnarray}
which amounts to a Gaussian convolution of the input plus rescaling of the phase space by $\sqrt{G}$.

Equation~(\ref{eqn:P-trans}) establishes two important points about PILA:\par\noindent
{\bf (i)} In general, a single-mode state can be represented by an $s$-parametrized phase-space distribution, $W_{s}(\alpha)$, with the parameter $s$ in the range $-1\le s\le1$---{\it e.g.},  Glauber-Sudarshan $P$-function ($s=1$), Wigner function ($s=0$), or $Q$-function ($s=-1$). This distribution is known to be well-behaved, with no singularity, for $s\le0$.
In addition, two different distributions are related to each other by a convolution \cite{Barnett}:
\begin{eqnarray}
W_{s^\prime}(\alpha)=\frac{2}{\pi(s-s')}\int d^2\beta W_{s}(\beta)\exp\mkern-2mu\left({-\frac{2}{(s-s')}|\alpha-\beta|^2}\right),
\label{eqn:dist}
\end{eqnarray}
where $s'<s$. Therefore we see, comparing Eqs.~(\ref{eqn:P-trans}) and (\ref{eqn:dist}), that the $P$-function always exists in a well-behaved form after amplification with $G\ge2$.\par\noindent
{\bf (ii)} Equation (\ref{eqn:P-trans}) may be connected to the nonclassical depth defined by C. T. Lee \cite{Lee1}.
The Glauber-$P$ function becomes positive when smoothed by a Gaussian of sufficient width,
\begin{eqnarray}
P^\prime(\alpha)=\frac{1}{\pi\tau}\int d^2\beta P(\beta)\exp\left(-\frac{1}{\tau}|\beta-\alpha|^2\right).
\label{eqn:Lee}
\end{eqnarray}
The minimum $\tau\in[0,1]$ required for a positive $P$-function is taken as a nonclassicality measure. Thus, by comparing Eqs.~(\ref{eqn:P-trans}) and (\ref{eqn:Lee}), we see that a state of nonclassical depth $\tau$ remains nonclassical under PILA up to the critical gain $G_c=1/(1-\tau)$. This shows that Gaussian (squeezed) states, which have nonclassical depth in the range $\tau\le1/2$, cannot survive the cloning limit, {\it i.e.}, for them $G_c\le2$.

In contrast to Gaussian states, ``maximally'' nonclassical states ($\tau=1$) remain nonclassical at any gain, {\it i.e.}, $G_c=1/(1-\tau)\rightarrow\infty$. It is therefore interesting to identify those states having $\tau=1$. There exists a sufficient, though not necessary, criterion to detect maximally nonclassical states \cite{Barnett1}: 
if $\rho$ is orthogonal to a certain coherent state---{\it i.e.}, if
$\langle\alpha|\rho|\alpha\rangle=0$ for some $\alpha$---the state
is maximally nonclassical. For instance, L{\" u}tkenhaus and Barnett
\cite{Barnett1} showed that all {\it pure} single-mode states, apart
from Gaussian states, satisfy this condition. We thus conclude that
all pure non-Gaussian states remain nonclassical under PILA at any
gain $G$. Furthermore, any state (pure or mixed) can be transformed
so that it becomes maximally nonclassical, and thus its
nonclassicality persists to arbitrarily large gain; this is achieved
by the photon-addition scheme, $\rho^\prime\sim \hat{a}^\dag\rho
\hat{a}$ \cite{Agarwal00}; $\rho^\prime$ necessarily satisfies
$\langle0|\rho'|0\rangle=0$.

\subsection{Observable nonclassical effects}
In the above we have shown that there is a broad class of
single-mode nonclassical states whose nonclassicality survives even
at arbitrarily large finite gain. Let us now ask which nonclassical
effects can be experimentally observed at high gain. If $\rho$ is
classical, its $P$-function is positive definite, so that we may
construct, for any operator $\hat f=f(\hat{a}^\dag,\hat{a})$, the
positive quantity
\begin{eqnarray}
\langle\mkern2mu:\mkern-3mu{\hat f}^\dag {\hat f}\mkern-3mu:\mkern2mu\rangle=\int d^2\alpha |f(\alpha^*,\alpha)|^2P(\alpha)>0,
\label{eqn:classical}
\end{eqnarray}
where $:\mkern3mu:$ denotes normal-ordering of the operators $\hat{a}^\dag$ and $\hat{a}$. A violation of this inequality is an indicator of nonclassicality. We aim, then, to construct an operator ${\hat T}={\hat f}^\dag {\hat f}$ that violates the inequality (\ref{eqn:classical}) for a state $\rho_G$ amplified at gain $G$.

Since the $P$-function is well-behaved for $G\ge2$, whenever $\rho_G$ is nonclassical at higher gain there must exist a {\it finite} region of phase-space (at least one), centered on a point $\alpha_0$, within which $P_G(\alpha)$ is negative. One can then readily construct a Gaussian function $T(\alpha)=\exp(-|\alpha-\alpha_0|^2/\sigma)$, with variance $\sigma$, such that $\int d^2\alpha P_G(\alpha)T(\alpha)<0$. 
In operator form the negativity is expressed as
\begin{eqnarray}
\langle\mkern2mu:\mkern-3mu{\hat T}\mkern-4mu:\mkern2mu\rangle=\langle[(\sigma-1)/\sigma]^{({\hat a}^\dag-\alpha_0^*)({\hat a}-\alpha_0)}\rangle<0,
\end{eqnarray}
where we use the normal-ordering relation $\exp(\theta\hat b^\dag \hat b)=\mkern4mu:\mkern-3mu\exp[(e^{\theta}-1)\hat b^\dag\hat b]\mkern-3mu:$ for a field operator $\hat b$ \cite{Barnett}.
It serves as an observable nonclassical effect that persists to large gain. For simplicity, we may set $\alpha_0=0$ (equivalently, displace the state by $-\alpha_0$), in which case the defined nonclassical effect is a property of the photon-number distribution, $P(n)$, and expressed by $\langle[(\sigma-1)/\sigma]^{{\hat a}^\dag{\hat a}}\rangle=\sum_{n=0}^\infty [(\sigma-1)/\sigma]^n P(n)<0$. It is testable in a photon-counting experiment.

As an example, the  one-photon Fock state, after amplification under PILA at gain $G$, has $P$-function
\begin{equation}
P(\alpha)=\frac1{\pi^2(G-1)^2}\left(\frac{|\alpha|^2}{1-1/G}-1\right)\exp\left[-|\alpha|^2/(G-1)\right],
\end{equation}
which is negative for $|\alpha|^2<1-1/G$. Its nonclassicality is observed as $\langle[(\sigma-1)/\sigma]^{{\hat a}^\dag{\hat a}}\rangle<0$, for any $\sigma<1$ and any $G$. In particular, by choosing $\sigma=1/2$ this becomes a condition on the photon-number parity---$\langle(-1)^{{\hat a}^\dag{\hat a}}\rangle<0$.

Alternatively, one can construct a closely related nonclassical effect, more accessible in practice, that is insensitive to detector efficiency. We adopt a test operator in phase-insensitive form, ${\hat f}=\sum_j c_j\hat a^{\dag j}\hat a^j$. A classical state must then satisfy Eq.~(\ref{eqn:classical}) for arbitrary complex coefficients $c_j$, which yields the requirement ${\rm Det}[M^{(n)}]\ge0$, for $n=1,2,\dots$ \cite{Agarwal,Agarwal0},
where $M^{(n)}$ is an $n\times n$ matrix whose elements,
\begin{eqnarray}
M_{ij}^{(n)}=\langle\mkern2mu:\mkern-3mu(\hat a^\dag\hat a)^{i+j-2}\mkern-3mu:\mkern2mu\rangle,
\end{eqnarray}
can be measured in photon-coincidence detection; ${\rm Det}[M^{(n)}]<0$, for any $n$, indicates nonclassicality---{\it e.g.}, with $n=2$, ${\rm Det}[M^{(2)}]<0$ indicates sub-Poissonian photon counts.

Photodetection with efficiency $\eta<1$ can be modeled by ideal
detection after the signal mode, $\hat{a}$, has been mixed with a
vacuum mode, $\hat{v}$, at a beam-splitter of transmissivity
$\sqrt{\eta}$. The detected mode, $\hat{a}_{d}$, is expressed by
$\hat{a}_{d}=\sqrt{\eta}\hat{a}+\sqrt{1-\eta}\hat{v}$, which gives
the matrix elements
$M_{ij,\rm{det}}^{(n)}=\eta^{i+j-2}\langle\mkern2mu:\mkern-3mu(\hat
a^\dag\hat a)^{i+j-2}\mkern-3mu:\mkern2mu\rangle$. The determinant
of this matrix turns out to depend on detection efficiency through
an overall scale factor only---it is proportional to
$\eta^{n^2-n}$---so the proposed nonclassical condition is formally
(though not in the magnitude of the violation relative to
experimental noises) insensitive to $\eta$ . One can show that the
condition ${\rm Det}[M^{(n)}]<0$ holds up to a critical gain
$G_c^{(n)}$, which can take a value far beyond $G=2$. We plot
$G_c^{(n)}$ for an input one-photon Fock state in Fig.~1; the
critical gain increases monotonically with $n$, which shows that
higher-order nonclassicality survives a larger gain.

\begin{figure}
\centerline{\scalebox{0.8}{\includegraphics{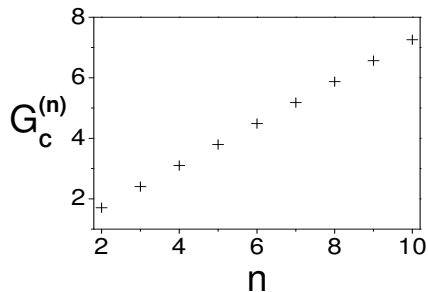}}}
\caption{ Critical gain $G_c^{(n)}$ at which the nonclassicality condition ${\rm Det}[M^{(n)}]<0$ ceases to be valid for the one-photon Fock state under PILA. }
\label{fig:fig1}
\end{figure}

\section{PILA-based quantum cloning}
We now turn our attention to the cloning of continuous variable (CV) states using PILA.
A quantum cloner based on PILA followed by beam-splitting is identified as the optimal Gaussian cloner for coherent states \cite{Cerf1,Cerf2} and has been experimentally implemented \cite{Andersen}. It may not be possible to construct a universal {\it optimal} cloner for the entire set of CV states, including all non-Gaussian states. Nevertheless, it does seem meaningful to take the PILA-based cloner as a universal machine for a broader set than the Gaussian states alone.
Suppose one amplifies a given input at gain $G$ and injects the amplified output into a series of $G-1$ beam splitters, each with transmissivity $t_i=1/\sqrt{G+1-i}$, $i=1,\dots,G-1$. In this configuration all the $G$ output clones are reduced to an identical local state having the same mean amplitude as the input, $\langle \hat{a}_{\rm clone}\rangle=\langle \hat{a}_{\rm in}\rangle$. In such symmetric cloning, one readily sees that the $P$-function of each clone is directly obtained from that of the amplified state---phase-space scaling by $1/\sqrt{G}$ due to beam-splitting---as $P_{\rm clone}(\alpha)\sim P_G({\sqrt G}\alpha)$. That is, using Eq.~(\ref{eqn:P-trans}),
\begin{eqnarray}
P_{\rm clone}(\alpha)=\frac{G}{\pi (G-1)}\int d^2\beta P_{\rm in}(\beta)\exp\left(-\frac1{1-1/G}|\beta-\alpha|^2\right),
\label{eqn:clone-P}
\end{eqnarray}
which, in view of Eq.~(\ref{eqn:dist}), is just the $s$-parametrized distribution of the input with $s=2/G-1>-1$.
We now deduce the following:\par\noindent
{\bf (i)} In a classical measure-then-prepare strategy, where position and momentum are simultaneously measured to estimate an unknown state, the $P$-function at the output equals the $Q$-function at the input \cite{Hammerer}.
One can thus see that the PILA-based quantum cloner is always superior, in terms of fidelity, to the classical scheme, regardless of the set of input states and number, $G$, of clones. The fidelity, $F$, between a pair of states, $\rho_1$ and $\rho_2$, is given by $F={\rm Tr}\{\rho_1\rho_2\}=\pi^{-1}\int d^2\lambda C_1(\lambda)C_2(-\lambda)$, where $C_i(\lambda)\equiv {\rm Tr}\{D_i(\lambda)\rho_i\}$ is the state characteristic function, with $D_i(\alpha)\equiv\exp(\alpha \hat{a}_i^\dag-\alpha^*\hat{a}_i)$, $i=1,2$, the displacement operator \cite{Barnett}. When $C_2(\lambda)$ is the same as the $s$-parametrized characteristic function of $\rho_1$---{\it i.e.}, $C_2(\lambda)=C_1(\lambda)\exp(s|\lambda|^2/2)$---the
fidelity is reduced to $F=\pi^{-1}\int d^2\lambda |C_1(\lambda)|^2\exp(s|\lambda|^2/2)$, which monotonically increases with $s$.
Therefore, the fidelity between each clone and the input state decreases with increasing number, $G$, of clones, owing to the fact that $s=2/G-1$; it is, however, higher than that of the classical scheme with $s=-1$.
 \par\noindent
{\bf (ii)} If the set of unknown input states has nonclassical depth
$\tau$, all clones remain nonclassical up to the critical clone
number $G_c=1/(1-\tau)$ [compare Eqs.~(\ref{eqn:Lee})
and~(\ref{eqn:clone-P})]. One cannot, for instance, have two clones
of a squeezed state remain squeezed, since $\tau\le1/2$ does not
give $G_c>2$. On the other hand, arbitrarily many clones of a state
from the set of all pure non-Gaussian states can be created as
nonclassical states since $\tau=1\Rightarrow G_c=\infty$.

\section{Two-mode case}
We now extend our treatment to the two-mode case. In particular, we want to see how a nonclassical correlation, {\it e.g.}, two-mode entanglement, evolves as each mode is amplified. An extension of Eq.~(\ref{eqn:P-trans}) gives the output two-mode $P$-function
\begin{eqnarray}
P_G(\alpha_1,\alpha_2)=\int d^2\beta_1 d^2\beta_2 P_{\rm in}(\beta_1,\beta_2)\prod_{i=1}^2Z_i(\alpha_i,\beta_i),
\label{eqn:P-trans-twomode}
\end{eqnarray}
where $Z_i=[\pi(G_i-1)]^{-1}\exp\mkern-3mu\left[-(1-1/G_i)^{-1}|\beta_i-\alpha_i/\sqrt{G_i}|^2\right]$, $i=1,2$. One readily sees, from Eqs.~(\ref{eqn:dist}) and~(\ref{eqn:P-trans-twomode}), that all Gaussian two-mode states, {\it e.g.}, squeezed states, become classical for $G\ge2$ as their original $s$-parametrized distributions are positive definite for $s\le0$ \cite{Agarwal1}. Thus, if entanglement is to survive above the cloning gain, $G=2$, its survival must be sought outside the Gaussian regime.

For clarity we consider a particular but broad class of entangled states; namely, those obtained by combining a single-mode nonclassical state, $\rho_{\rm in}$, and a vacuum state at a beam splitter of transmissivity $t=\cos\theta$ \cite{Asboth}. This class serves as a useful tool to study many interesting entangled states: if $|\Psi_{\rm in}\rangle$ is the Fock state $|N\rangle$, for example, one obtains a bimodal two-mode entangled state \cite{nha1}, the simplest being the single-photon state $\cos\theta|1,0\rangle+\sin\theta|0,1\rangle$; if $|\Psi_{\rm in}\rangle$ is the Schr\"odinger-cat state $|\alpha\rangle-|-\alpha\rangle$, one obtains the entangled coherent states $|\alpha\cos\theta,\alpha\sin\theta\rangle-|-\alpha\cos\theta,-\alpha\sin\theta\rangle$.

Let the input to the beam splitter be represented by the $P$-function $P(\alpha_1,\alpha_2)=P_{\rm in}(\alpha_1)\delta^{(2)}(\alpha_2)$.
Then, from the beam-splitter transformation, $\hat{a}_1^\prime=\hat{a}_1\cos\theta+\hat{a}_2\sin\theta$,
$\hat{a}_2^\prime=-\hat{a}_1\sin\theta+\hat{a}_2\cos\theta$, and  Eq.~(\ref{eqn:P-trans-twomode}), the amplified two-mode state has $P$-function
\begin{eqnarray}
P_G(\alpha_1,\alpha_2)={\cal N}_{12}\int d^2\beta P_{\rm in}(\beta)\exp\left(-{\cal G}_{12}\left|\beta-\alpha_{12}/{\cal G}_{12}\right|^2\right),
\label{eqn:P-global-twomode}
\end{eqnarray}
where
\begin{eqnarray}
{\cal N}_{12}&&\equiv \frac{{\rm exp}\left[|\alpha_{12}|^2/{\cal G}_{12}-|\alpha_1|^2/(G_1-1)-|\alpha_2|^2/(G_2-1)\right]}{\pi^2(G_1-1)(G_2-1)},\nonumber\\
\noalign{\vskip3pt}
{\cal G}_{12}&&\equiv\lambda_1^{-2}\cos^2\theta+\lambda_2^{-2}\sin^2\theta,\nonumber \\
\noalign{\vskip3pt}
\alpha_{12}&&\equiv\frac{1}{\lambda_1^2\sqrt{G_1}}\mkern3mu\alpha_1\cos\theta+\frac{1}{\lambda_2^2\sqrt{G_2}}\mkern3mu\alpha_2\sin\theta,\nonumber
\end{eqnarray}
with $\lambda_i\equiv\sqrt{1-1/G_i}$, $i=1,2$.
On the other hand, the local state of mode 1 has the $P$-function
\begin{eqnarray}
P_G(\alpha_1)=\frac{1}{\pi(G_1-1)}\int d^2\beta P_{\rm in}(\beta)\exp\mkern-3mu\left(-\frac{\cos^2\theta}{\lambda_1^2}\left|\beta-\alpha_{1}/\sqrt{G_1}\cos\theta\right|^2\right),
\label{eqn:P-singlemode}
\end{eqnarray}
with $\cos\theta\rightarrow\sin\theta$ in the similar expression for mode 2.
Equations~(\ref{eqn:P-global-twomode}) and~(\ref{eqn:P-singlemode}) show how single- and two-mode nonclassicality decohere under PILA.
Let us consider, in particular, symmetric PILA, {\it i.e.},  $G_1=G_2=G$. Then, if the input to the beam splitter, $P$-function $P_{\rm in}(\alpha_1)$, has nonclassical depth $\tau$,
one sees from Eqs.~(\ref{eqn:Lee}) and~(\ref{eqn:P-global-twomode}) that the global two-mode state remains nonclassical below the critical gain $G_c=1/(1-\tau)$. On the other hand, from Eqs.~(\ref{eqn:Lee}) and~(\ref{eqn:P-singlemode}), the local states remain nonclassical up to $G_{c1}=1/(1-\tau\cos^2\theta)$ and $G_{c2}=1/(1-\tau\sin^2\theta)$, respectively. The local critical values are smaller than the global critical value, $G_{c1,2}<G_c$; thus, for the range of gain satisfying $G_{c1,2}\le G<G_c$, 
the nonclassicality of the amplified state exists in the form of {\it correlation\/} exclusively.

To take an example, when the entangled state $|\alpha,\alpha\rangle-|-\alpha,-\alpha\rangle$ or $|1,0\rangle+|0,1\rangle$ is amplified ($\tau=1$),
this analysis setting $\theta=\pi/4$ shows that the nonclassical correlation survives at arbitrarily large gain, with $G_c=\infty$, whereas all local nonclassicality disappears above the cloning limit, $G=2$. Since the amplified fields possess a well-behaved $P$-function at high gain, the method of \cite{Kiesel} might be extended to the two-mode case to directly measure the $P$-function, thus demonstrating persistent nonclassical correlation for experimentally accessible non-Gaussian states.

\subsection{Persistent entanglement}
We must ask whether the surviving nonclassical correlation is indeed attributable to entanglement, since Marek {\it et al.}~\cite{Marek} find that separable states also can exhibit nonclassical correlation. It can be shown that the non-Gaussian states considered here are negative under partial transposition, a direct proof that they are entangled \cite{Peres}. Let us illustrate persistent entanglement for the class of states generated by combining, at a 50:50 beam splitter, a maximally nonclassical single-mode state satisfying the condition $\langle \alpha|\rho_{\rm in}|\alpha\rangle=0$ \cite{Barnett1} and a vacuum state. We prove that all such two-mode states remain entangled for any gain $G$. We may, without loss of generality, set $\alpha=0$; local displacement before the beam splitter is equivalent to local displacement after amplification, which does not change the entanglement. Thus, the input-state condition can be taken as $\langle 0|\rho_{\rm in}|0\rangle=0$, {\it i.e.}, vanishing vacuum-state probability. We denote by $N$ the lowest nonzero occupation number of $\rho_{\rm in}$.

Note now that the actions of the beam-splitter and PILA commute in symmetric amplification, as one readily verifies in the Heisenberg picture.
Furthermore, the condition $\langle j|\rho_{\rm in}|j\rangle=0$, $j<N$, holds also for the amplified state, $\rho_{{\rm in},G}$, which can be  confirmed from Eq.~(\ref{eqn:Milburn}).  On the other hand, after amplification at gain $G$ a vacuum state becomes a thermal state, $|0\rangle\langle0|_G\equiv G^{-1}\sum_n(1-1/G)^n|n\rangle\langle n|$.
Therefore, for the two-mode amplified state, $\rho_{E,G}=U_{\rm BS}\cdot\rho_{{\rm in},G}\otimes|0\rangle\langle0|_G\cdot U_{\rm BS}^\dag$, we obtain
\begin{eqnarray}
\langle 0,j|\rho_{E,G}|j,0\rangle&&=\frac{1}{j!}\langle 0,0|\hat{a}_2^j\rho_{E,G}\hat{a}_1^{\dag j}|0,0\rangle\nonumber\\
&&=\frac{1}{2^jj!}\langle 0,0|(-\hat{a}_1+\hat{a}_2)^j\cdot\rho_{{\rm in},G}\otimes|0\rangle\langle0|_G\cdot (\hat{a}_1^\dag+\hat{a}_2^\dag)^j|0,0\rangle\nonumber\\
&&=\frac{(-1)^j}{2^jG}\langle j|\rho_{{\rm in},G}|
j\rangle.\hspace{2cm} (j\le N) \label{eqn:element}
\end{eqnarray}
Equation~(\ref{eqn:element}) gives a zero diagonal element $\langle0,0|\rho_{E,G}|0,0\rangle=0$ and nonzero off-diagonal element $\langle 0,N|\rho_{E,G}|N,0\rangle\ne0$.
The entanglement of $\rho_{E,G}$ is then proved by negativity under partial transposition in the subspace spanned by $|0,0\rangle$ and $|N,N\rangle$;
the determinant of $\rho_{E,G}$ under partial transposition is $\langle0,0|\rho_{E,G}|0,0\rangle\langle N,N|\rho_{E,G}|N,N\rangle-\langle 0,N|\rho_{E,G}|N,0\rangle^2<0$.

To further verify this persistent entanglement, known methods based on partial transposition can be used  \cite{criterion}---{\it e.g.}, for amplification of the state $|1,0\rangle+|0,1\rangle$, a witness operator $W=|e_-\rangle\langle e_-|^{\rm PT}$, with $|e_-\rangle=|0,0\rangle-(\sqrt{1+\lambda^4}-\lambda^2)|1,1\rangle$ and $\lambda=\sqrt{1-1/G}$,
is constructed to show that ${\rm Tr}\{W\rho_{E,G}\}<0$.

\subsection{Cloning of quantum correlation}
Finally, we briefly discuss the case where PILA followed by beam
splitting is applied as in Sec.~III to each of two modes, to achieve
quantum cloning of nonclassical correlation or entanglement. Our
analysis shows that if the PILA-based quantum cloner makes multiple
copies of a two-mode entangled state, any pair of clones possesses a
nonclassical correlation up to the critical clone number
$G_c=1/(1-\tau)$; the beam splitter following PILA simply rescales
the $P$-function, leaving its negativity unchanged. On the other
hand, negativity under partial transposition disappears at the
two-clone limit ($G=2$) for symmetric cloning, due to vacuum noise
at the beam-splitter. A subtle contrast between nonclassical
correlation and entanglement therefore emerges in the symmetric
quantum cloning of two-mode non-Gaussian entangled states
\cite{Marek}.

If, however, only one party makes clones ($G_2=0$), it turns out
that entanglement survives between any clone of mode 1 and the
original copy of mode 2 regardless of $G_1$ for both Gaussian and
non-Gaussian pure inputs \cite{Lett}. This can be explained in a
general framework of one-sided quantum channels for CVs and more
details will be investigated elsewhere.

\section{Concluding remarks}
We have shown that certain nonclassical effects persist far beyond
the cloning limit (at gain $G=2$) in quantum-limited PILA. We
identified a broad class of single-mode and two-mode non-Gaussian
states that survive arbitrarily large gain and associated observable
effects that are testable by photon counting. Remarkably, these
persistent nonclassical effects include quantum entanglement for
two-mode non-Gaussian fields. This result is to be contrasted with
that of Ref. \cite{Paris}, which provided some evidence for the
maximal robustness of Gaussian entanglement at fixed energy in the
presence of thermal noise. Although it was further conjectured that
Gaussian entangled states would generally be maximally robust under
noisy Markovian interactions, our work provides a counter-argument;
Gaussian entanglement disappears above $G$=2, whereas non-Gaussian
entanglement persists beyond $G=2$ under the Markovian evolution of
PILA \cite{note}. Our results also show that a quantum cloner based
on PILA can produce many nonclassical clones with better fidelity
than the classical measure-then-prepare scheme for a set of
non-Gaussian states. Demonstration of these results seems feasible
in view of recent experimental progress
\cite{Josse,Andersen,Kiesel}.

\section*{Acknowledgement}
HN is supported by the NPRP grant 1-7-7-6 from the Qatar National Research Fund,  GJM by the Australian Research Council Center of Excellence in Quantum Computer Technology, and HJC by the Marsden fund of RSNZ.


\begin{thebibliography}{10}
\bibitem{note} The PILA master equation reads as
$\dot\rho=\gamma N_2(2a^\dag\rho a-aa^\dag\rho-\rho aa^\dag)+\gamma N_1(2a\rho a^{\dag}-a^\dag a\rho-\rho a^\dag a)$, with $N_1$ and $N_2$ being the numbers of unexcited and excited atoms, respectively. The quantum-limited amplifier considered here is characterized by $N_1=0$ and gain $G=e^{2\gamma N_2 t}$.
\end{thebibliography}
\end{document}